\documentclass[reprint,superscriptaddress,amsmath,amssymb,prl]{revtex4-2}

\usepackage{graphicx}
\usepackage{dcolumn}
\usepackage{bm}
\usepackage[mathlines]{lineno}

\usepackage {mathtools} 
\usepackage{float}

\usepackage[colorlinks]{hyperref}

\def\be{\begin{equation}}
\def\ee{\end{equation}}
\def\bea{\begin{eqnarray}}
\def\eea{\end{eqnarray}}

\begin{document}

\title{Two dimensional arrays of Bose-Einstein condensates: interference and stochastic collapse dynamics }

\author{Youjia Huang}
\affiliation{Department of Physics and State Key Laboratory of Low Dimensional Quantum Physics, Tsinghua University, Beijing, 100084, China}

\author{Shu Nagata}
\affiliation{James Franck Institute, Enrico Fermi Institute, and Department of Physics, University of Chicago, Chicago, Illinois 60637, USA}

\author{Joseph Jachinowski}
\affiliation{James Franck Institute, Enrico Fermi Institute, and Department of Physics, University of Chicago, Chicago, Illinois 60637, USA}

\author{Jiazhong Hu}
\email{hujiazhong01@ultracold.cn}
\affiliation{Department of Physics and State Key Laboratory of Low Dimensional Quantum Physics, Tsinghua University, Beijing, 100084, China}
\affiliation{Frontier Science Center for Quantum Information and Collaborative Innovation Center of Quantum Matter, Beijing, 100084, China}

\author{Cheng Chin}
\email{cchin@uchicago.edu}
\affiliation{James Franck Institute, Enrico Fermi Institute, and Department of Physics, University of Chicago, Chicago, Illinois 60637, USA}

\date{\today}

\begin{abstract}
We demonstrate two-dimensional arrays of Bose-Einstein condensates (BECs) as a new experimental platform with parallel quantum simulation capability. A defect-free array of up to 49 BECs is formed by loading a single BEC with 50,000 atoms into 7$\times$7 optical wells. Each BEC is prepared with independent phases, confirmed by matterwave interference. Based on BEC arrays, we realize fast determination of the phase boundary of BECs with attractive interactions. We also observe the stochastic collapse dynamics from the distribution of atom numbers in the array. We show that the collapse of a BEC can occur much faster than the averaged decay of an ensemble. The BEC arrays enable new forms of experiments to drastically increase the measurement throughput and to quantum simulate, say, large 2D Josephson-junction arrays. 

\end{abstract}

\maketitle

Optical tweezer arrays have emerged as a new exciting platform to realize independent control of individual atoms, enabling experiments on atomic qubits for applications in quantum information and quantum metrology \cite{kaufman2021quantum,toth2014quantum,zhang2016precision,riedel2010atom,pezze2018quantum}. This contrasts the optical lattice platform, which does not offer easy control over individual atoms, but has the advantage of keeping a large number of atoms in the quantum degenerate regime \cite{bloch2008many}.

To combine the strengths of both platforms, innovative ideas have been investigated to employ the concept of optical tweezer to manipulate multiple quantum degenerate ensembles \cite{wang2020preparation,trisnadi2022design,young2022tweezer}. One approach is to prepare an array of Bose-Einstein condensates (BECs) by loading a single BEC into multiple optical potential wells. Based on a few BECs in a 1D chain, quantum simulation of Josephson-junctions has been realized \cite{cataliotti2001josephson,gati2006realization,gati2007bosonic,anderson1998macroscopic}. Quantum simulation of 2D Josephson-junction array has also been proposed based on 2D array of BECs~\cite{kopec2013berezinskii}. In addition, each BEC in the array is envisaged as a quantum memory unit \cite{duan2001long,colombe2007strong,riedl2012bose}. Finally since the cycle of the BEC experiment is typically long, preparation of many BECs in an array can significantly improve the experimental throughput and statistics.  

In this paper, we demonstrate a two-dimensional (2D) array of up to 7$\times$7 BECs in a single experiment. This is realized by adiabatically loading a single atomic BEC into a 2D array of optical wells, each supporting a small BEC. By interfering the BECs in the array, we show that the BECs after the transfer possess independent phases. We use the 2D array to realize parallel experiments to quickly determine the stability phase diagram of BECs with attractive interactions. By directly comparing BECs in the array, we show that the collapse dynamics of a BEC is stochastic in nature.

\begin{figure}[htbp]
\centering
\includegraphics[width=0.48\textwidth]{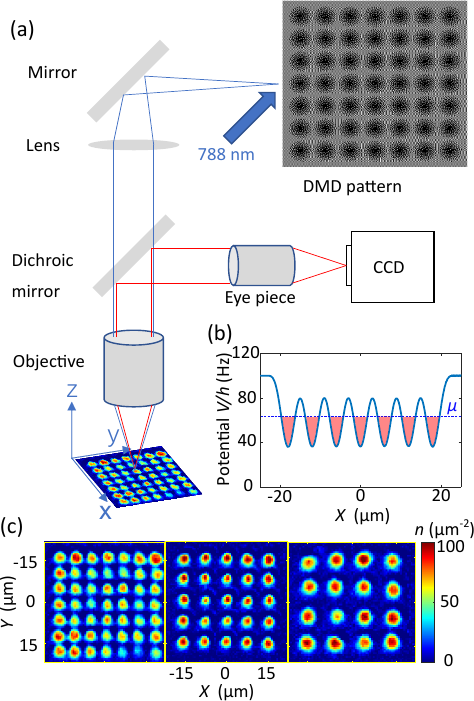}
\caption{
Experimental preparation of two-dimensional (2D) arrays of BECs
(a) A digital micro-mirror device (DMD) is placed at the image plane of the atomic sample. A 788~nm beam illuminates the DMD, which projects the desired intensity pattern on the atoms. A pattern based on $341 \times 341$ micromirrors is shown, which generates $7 \times 7$ optical wells. Here bright pixels correspond to the micromirrors that reflect the light to the atoms; dark pixels correspond to the micromirrors that do not. (b) A 1D line cut of 7 optical potential wells on the atom plane. We raise the barriers between the wells higher than the chemical potential $\mu$ to separate individual BECs.
(c) Images of 2D array with $7\times 7$, $5\times 5$, and $4\times 4$ BECs. Each site contains 1,000 $\sim$2,000 atoms. The color represents the atomic density.
}
\label{fig:Fig_1}
\end{figure}

We start the experiment by preparing a BEC of $5\times10^4$ cesium atoms at the scattering length $a = 200 a_0$, where $a_0$ is the Bohr radius. In the vertical direction, the BEC is tightly confined to a single site of an optical lattice with trap freq $\omega_z = 2 \pi \times 400$~Hz \cite{Supplement}.
In the horizontal directions, the BEC is weakly confined by a flat-bottomed square potential well with a side length of 50~$\mu$m and barrier height $h\times$ 370~Hz, where $h = 2\pi\hbar$ is the Planck constant. The optical potential is formed by projecting a blue-detuned light at 788~nm with a digital micromirror device (DMD), see Fig.~\ref{fig:Fig_1}a and details in Ref.~\cite{Supplement}. 

We divide the initial BEC into an array of $M^2$ small BECs by adiabatically loading the BEC into an $M\times M$ array of potential wells formed by the DMD. The separation between adjacent wells is $d=$ 11.7, 8.4, 5.8~$\mu$m for $M=$ 4, 5 and 7, respectively. The depth of the potential well is $h\times20$ to $h\times80$~Hz, calculated from the trap frequency measurement. To ensure adiabaticity, the blue-detuned light is slowly turned on in 200~ms and the scattering length is simultaneously ramped to a small final value of $a_f<$20~$a_0$ using a Feshbach resonance \cite{chin2010feshbach}. This procedure reduces the chemical potential and isolates each BEC in a single well, see Fig.~\ref{fig:Fig_1}b. After loading, we perform \textit{in situ} or time-of-flight (TOF) imaging on the atoms to confirm the preparation process, see examples in Fig.~\ref{fig:Fig_1}c. 

\begin{figure}[htbp]
\centering
\includegraphics[width=0.48\textwidth]{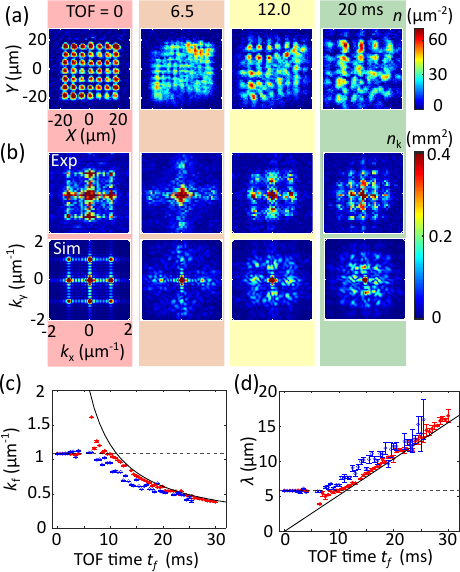}
\caption{
Interference of a two-dimensional array of BECs
(a) Images of 49 BECs during the time-of-flight expansion. Here atomic scattering length is $a_f$ = 3~$a_0$ (b) Fourier transform of the atomic density profile (upper). Numerical simulation under the same condition (lower). (c) The wave vector $k_f$ of the main side peaks in the Fourier spectrum. Red circles are data taken at $a_f$ = 3~$a_0$; Blue circles are data taken at $a_f$ = 75~$a_0$. The dash line indicates the initial BEC separation of $5.8$~$\mu$m. The solid line indicates the predicted interference wavevector $k_f=md/\hbar t_f$.
(d) The period of density modulation extracted from the Fourier transform. The dash line is the separation between optical wells. The solid line is the predicted spacing $\lambda=ht_f/md$. Error bars show one standard deviation. 
}
\label{fig:inter}
\end{figure}

To verify that the initial BEC is converted into $M^2$ separate BECs, we interfere them from different wells in a TOF experiment. Interference of two BECs \cite{andrews1997observation,rohrl1997transition,shin2004atom} and 1D BEC array \cite{hadzibabic2004interference,gustavsson2010interference,pedri2001expansion} were studied previously. 
The fringe period due to two interfering BECs is $\lambda=h t_f/md$ where $t_f$ is the TOF time and $m$ is the atomic mass.The same fringe spacing is observed for 1D BEC arrays with random phases \cite{hadzibabic2004interference}.

We implement the interference experiment with a BEC loaded into $7\times 7$ optical wells. We quickly remove all horizontal confinement, which allows the atoms from different wells to freely expand, see Fig.~\ref{fig:inter}. Clear density waves appear when the expanding BECs overlap, see Fig.~\ref{fig:inter}a. To determine the periodicity of the waves, we Fourier transform the atomic density distribution and extract the dominant non-zero wavenumber $k_f$ in the momentum space \cite{Supplement}. The result and the associated period $\lambda=2\pi/k_f$ are shown in Figs.~\ref{fig:inter}b-d.

For short expansion times $t_f\leq$5~ms, there is insufficient overlap of BECs from different wells. Thus the period of the density variation is simply given by the separation of optical wells $d=$5.8~$\mu$m. For $t_f > 6$~ms, the interference pattern emerges with a spatial period $\lambda$ which increases with time, consistent with the theoretical prediction. Repeating the experiment with the atomic scattering length tuned to $a_f = 3$ and 75~$a_0$ right before the TOF, we observe a slightly longer period for 75~$a_0$. 

The appearance of matter interference confirms the preparation of a 2D BEC array. We further investigate the phase coherence of the BECs in different wells by comparing our experimental results to numerical simulation of the Gross-Pitaevskii equation (GPE). In the simulation we find that if the phases of the BECs are identical, the potential well spacing $d$ persists as the dominant length scale over hundreds of ms; in contrast, if the phases of the BECs are random, the initial periodicity $d$ is rapidly replaced by the interference fringe spacing \cite{Supplement}. In our experiment, the initial periodicity $d$ only persists in the first 5~ms, and the interference fringes spacing become dominant afterwards. Our result is fully consistent with independent BECs prepared with random matterwave phases.

An intriguing application of BEC arrays is to perform parallel experiments. One experimental cycle on a $M \times M$ BEC array yields $M^2$ measurements and thus increases the data throughput. In addition, BECs can be prepared in different conditions, which allow us to directly compare many systems under otherwise an identical environment. Such parallel experiments are less sensitive to systematics than the standard repetitive experiments. 

We demonstrate the power of parallel experiments with BEC arrays by investigating the stability phase boundary of BECs with negative scattering length $a<0$. Collapse of a BEC occurs when the atomic attraction or the atomic density exceeds a critical value. BEC collapse has been studied in former experiments \cite{bradley1997bose,sackett1998growth, duine2001stochastic, kagan1998collapse, perez1997dynamics,huepe1999decay, roberts2001controlled}. The stability condition is theoretically calculated to be $-a/\ell < 0.573/N$ \cite{houbiers1996stability}, where $\ell=\sqrt{\hbar/m\bar{\omega}}$ is the harmonic length of the trap and $\bar{\omega}$ is the geometric mean of the trap frequencies.

To determine the phase boundary, an array of $4\times 4$ BECs with initial scattering length 4~$a_0$ is employed. We increase the BEC separation to 8~$\mu$m and the potential barrier between them to 4~nk to ensure the independent evolution of each BEC. The scattering length is then quickly quenched to different negative values $a_f<0$. After a hold time $t_h$, we image the BECs and see if they have collapsed. 

\textit{In situ} imaging clearly distinguishes collapsed and stable BECs after a hold time $t_h = 50$~ms, see  Fig.~\ref{fig:collapse}a. In stable BECs, the atom number remains essentially unchanged, while the atom number drops sharply to 20$\sim$40$\%$ in the collapsed BECs. The dichotomy of stable and collapsed BEC manifests in the histogram of the remaining atom number, see Fig.~\ref{fig:collapse}b. The histogram shows two distinct peaks at 100$\%$ and 35$\%$ of the initial number. In very few cases do we see BECs with an atom number falling between 60$\sim$80$\%$. Thus we introduce the threshold fraction of 70$\%$, below (above) which the BEC is considered collapsed (stable). 

The 16 BECs in the array allow us to efficiently determine the phase boundary between stable and collapsed BECs for various initial particle numbers and scattering lengths. We start the experiment with an array of $4 \times4 $ BECs, where each BEC is prepared with a different atom number between 1,000 and 2,000 by slight tuning of the trap depth of each well. We verify that the initial population is reproducible to better than $90\%$. The non-uniform preparation of the BECs tests the dependence of the BEC stability on the atom number. Repeating the experiment for different scattering lengths $a_f$, we see a clear boundary between stable and unstable BECs. Our result shows that the critical scattering length $a_c$ reduces for BECs with more atoms, consistent with the theoretical prediction $a_c / \ell = -\kappa /N$, where $\kappa = 0.573$ \cite{houbiers1996stability}.  An independent fit to our data yields $\kappa = 0.51(7)$, see Fig.~\ref{fig:collapse}. Our result is consistent with the former measurement of $\kappa=0.46(6)$ \cite{roberts2001controlled} and the theoretical prediction within our measurement uncertainty.

\begin{figure}[htbp]
\centering
\includegraphics[width=0.48\textwidth]{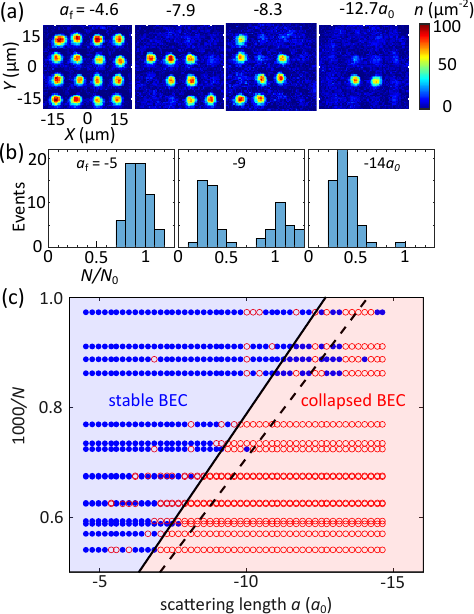}
\caption{Collapse of BECs in a 2D array with attractive interactions.
(a) Images of BEC arrays after a hold time of $t_h=50$~ms. The atomic population quickly drops when BEC collapses. The trap frequencies approximately $(\omega_x, \omega_y, \omega_z)=2\pi\times$(15,15,400)~Hz. 
(b) Histograms of remaining atom number $N$ normalized to the atom number $N_0$ at $a_f=$~-4.6~$a_0$. (c) Stability phase diagram of BEC with attractions constructed based on 2D arrays of BECs with different initial atom numbers $N$ and scattering lengths $a_f$. Blue filled circles are stable BECs and red open circles are collapsed BECs. Dashed line shows the predicted critical scattering length $a_c= -0.574 \ell/N$. Solid line is a fit to the data which yields $a_c=-0.51(7)\ell/N$. Details see in Ref.~\cite{Supplement}.
}
\label{fig:collapse}
\end{figure}

\begin{figure*}[htp]
\centering
\includegraphics[width=1\textwidth]
{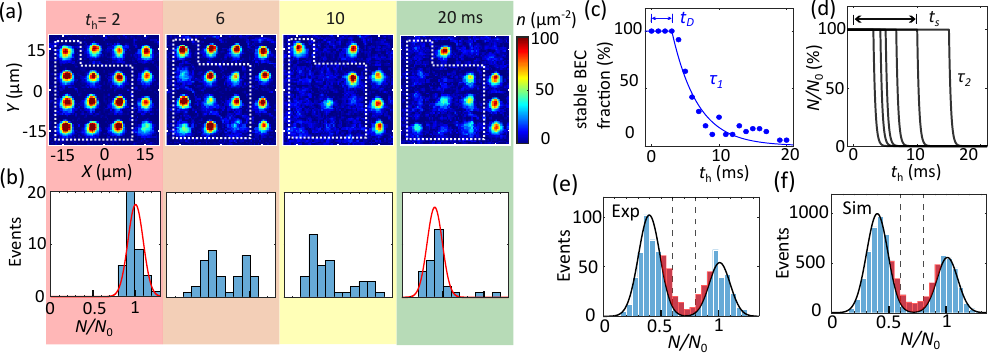}
\caption{Stochastic collapse of BECs with attractive interactions (a) An arrays of 16 BECs with scattering length quenched to $a_f=-9$~$a_0$. We build statistics based on samples with similar initial atom number (within $10\%$), enclosed by the white dashed line. (b) Histograms of normalized atom numbers $n\equiv N/N_0$ based on $4$ repeated experiments. Gaussian fit to the histogram at $t_h=2$~ms gives the mean $\bar{n}=1.0$ and standard deviation $\sigma=0.09$ for stable BECs, and the fit to histogram at $t_h=20$~ms gives $\bar{n}=0.4$ and $\sigma=0.09$ for the collapsed BECs. (c) Averaged BEC survival fraction remains unity for $t_D=3$~ms after the quench. Then decay occurs with a time constant of $\tau_1=3.8(9)$~ms. Blue line shows an empirical fit. (d) Illustration of our statistical model, where each BEC is stable for a stochastic time $t_s$, followed by the collapse that occurs within a $1/e$ time of $\tau_2$ (black lines). (e) Histogram of atom number of all 810 samples with 0$<t_d<$~20~ms hold time. Black lines are fits based on the stable and collapsed BEC distributions from panel (b). The red parts of the histogram are excess events of BECs with intermediate atom number 0.5$<\bar{n}<$1. (f) Histogram from the theory model based on 8,100 samples. By comparing data and model, we determine the collapse process of a single BEC takes $\tau_2$ = 1.3(2)~ms
}
\label{fig:Fig_3}
\end{figure*}

Near the boundary between stable and collapsed BECs, we observe unusually large fluctuations in the remaining atom number. This, together with the double-peak structure in the histogram, suggests that the collapse is likely a highly stochastic process. This scenario has been discussed in Ref.~\cite{sackett1998growth}. To better understand the collapse process, we prepare a 4$\times$4 BEC array with almost identical initial atom numbers at scattering length $-1$~$a_0$. We then quickly quench the scattering length near the phase boundary at $a_f = -9$~$a_0 = -0.5~\ell/N$ and monitor their evolution. We build statistics of the atom number based on 10 of the 16 sites whose initial atom numbers are stable and close to each other, see Figs.~\ref{fig:Fig_3}a and b. After $t_h = 4$~ms, BECs start collapsing and we again observe two peaks in the histogram, indicating some BECs remain stable and while some have fully collapsed. After 20~ms, almost all BECs have collapsed. Based on 810 samples, an ensemble average of the survival fractions shows that BECs remain stable for a delay time of $t_D = 3$~ms and then collapse occurs with an averaged $1/e$ lifetime of $\tau_1=3.8(9)$~ms, see Fig.~\ref{fig:Fig_3}c. 

During the collapse process, the two-peak structure in the histogram persists. This implies that the collapse occurs within a time scale much shorter than the ensemble-averaged decay time $\tau_1$. In other words, the collapse happens so quickly that few BECs are recorded during the process. 
This is analogous to the radiative decay of heavy elements, where the half-life can be much longer than the time scale of the nuclear reaction.

We construct a stochastic model to describe the decay of metastable BECs with attractive interactions. After quenching the magnetic field, we assume each BEC takes a duration of $t_D$ to reach the atomic density that is high enough to initiate the collapse process. This observation is consistent with previous works \cite{donley2001dynamics,nguyen2017formation} and is explained by the development of modulation instability \cite{nguyen2017formation}. The collapse then occurs stochastically with a time constant of $\tau_1$.

We introduce a new microscopic time scale $\tau_2$ in the model to characterize the duration of the collapse process of a single BEC. To extract $\tau_2$, we numerically calculate the histogram for different $\tau_2$ and compare with the measurement \cite{Supplement}. We focus on the events falling between the two peaks in the histogram and show that among 810 BEC samples in the 2D arrays, less than 5$\%$ of the events are between the two peaks that cannot be accounted for by the initial and collapsed BEC atom number distributions. These events are marked red in Figs.~\ref{fig:Fig_3}e and f. The scarcity of these events indicates that collapse in a BEC occurs so fast that one can hardly capture the BECs in the decay process. By comparing with the numerical model \cite{Supplement}, we determine the microscope collapse time scale to be $\tau_2=1.3(2)$~ms. 

The stable time $t_D=3$~ms and the decay time of $\tau_1=3.8$~ms suggest that the BECs can remain stable for approximately 7~ms. Once the collapse occurs, however, the BEC decays rapidly within 1.3~ms. This picture is illustrated in Fig.~\ref{fig:Fig_3}d. 

In the mean field picture, the metastability of BECs with small attraction comes from the quantum pressure that balances the attraction~\cite{pethick2008bose}. When the attraction approaches the critical value $a_f\rightarrow a_c$, however, stochastic collapse of the BECs can occur due to the suppressed kinetic energy barrier, which can be overcome by thermal or quantum fluctuations of the BECs. Our picture also explains the observation in the $^{85}$Rb BEC experiment \cite{donley2001dynamics}, where the averaged collapse dynamics is found to be much slower than the theory expectation \cite{wuster2007quantum}. This is because averaged lifetime of metastable BECs can be much longer than the microscopic time scale to collapse a single BEC.

In summary, we demonstrate 3 forms of experiments with 2D BEC arrays. First, matterwave interference of 49 BECs shows that the each BEC acquires an independent phase. Second, when each BEC is prepared differently, we realize fast mapping of the phase diagram. Finally, parallel experiments on arrays of BECs with attractive interactions reveal the stochastic nature of the collapse. Collapse of a single BEC can occur much faster than the ensemble-averaged decay. 

With sufficient atoms in a BEC, our scheme can be easily generalized to prepare 100 to 1,000 BECs in a single experiment. Such quantum gas array platform will invite more innovations in research on quantum simulation and quantum information processing. 

This work is supported by National Science Foundation under Grant No. PHY-1511696 and PHY-2103542. Y.~H. and J.~H. acknowledge support from the National Key Research and Development Program of China (2021YFA0718303, 2021YFA1400904) and National Natural Science Foundation of China (92165203, 61975092, 11974202). S.N. acknowledges support from the Takenaka Scholarship Foundation.

\bibliography{manuscript}

\clearpage
 \renewcommand\thefigure{S\arabic{figure}}

\setcounter{figure}{0}   

\maketitle

\onecolumngrid
\begin{center}

\begin{large}
\textbf{Supplementary Material for\\Two dimensional arrays of Bose-Einstein condensates: interference and stochastic collapse dynamics}
\end{large}\\

Youjia Huang$^{1}$, Shu Nagata$^{2}$, Joseph Jachinowski$^{2}$, Jiazhong Hu$^{1,3}$, and Cheng Chin$^{2}$\\
\emph{ $^{1}$Department of Physics and State Key Laboratory of Low Dimensional
Quantum Physics, Tsinghua University, Beijing, 100084, China\\$^{2}$James Franck institute, Enrico Fermi institute, and Department of Physics, University of Chicago, Chicago, Illinois 60637, USA\\$^{3}$Frontier Science Center for Quantum Information and Collaborative
Innovation Center of Quantum Matter, Beijing, 100084, China}
\vspace{1cm}
\twocolumngrid
\end{center}

\subsection*{A.Preparation of Two-Dimensional (2D) Array of Bose-Einstein Condensates (BECs)}

\begin{figure}[htbp]
\centering
\includegraphics[width=0.48\textwidth]{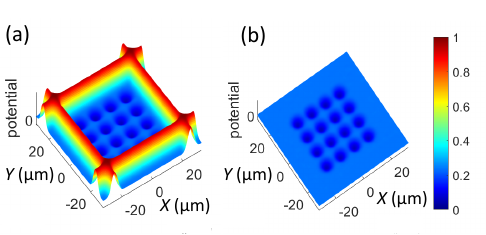}
\caption {Optical potentials for trapping initial single BEC (a) and an array of $M^2$ BECs (b). (a) A single BEC with high chemical potential is initially confined by four walls generated by the DMD. (b) After reducing the scattering length, the BEC is converted into an array of small BECs, confined by the optical wells}
\label{fig:Fig_S3}
\end{figure}

Our experiment begins by preparing approximately $5\times10^4$ atoms in a three-dimensional BEC with scattering length $a = $~200~$a_0$, where $a_0$ is the Bohr radius. Then we gradually turn on a  DMD potential with shallow 2D wells and repulsive square boundary in 200ms(Fig.~\ref{fig:Fig_S3}a). Next we turn on a vertical lattice to transfer atom to transfer the BEC into a single lattice site in 600~ms. At the same time, we ramp down the scattering length to $a = $~20~$a_0$ in 300~ms. As scattering length becomes lower, the chemical potential also drops and the atoms settle into the shallow wells. We then switch the DMD pattern to turn off the repulsive square boundary (Fig.~\ref{fig:Fig_S3}b) which allows the hot atoms to escape. Then we tune the scattering length $a$ to an even lower value (4~$a_0$ for experiments in Fig.~\ref{fig:collapse} and -1~$a_0$ for  experiments in Fig.~\ref{fig:Fig_3}) before proceeding to the experiments described in main text.

\subsection*{B. Determination of Fringe Period and Gross–Pitaevskii equation(GPE) Simulation}

\begin{figure}[!h]
\centering
\includegraphics[width=0.48\textwidth]{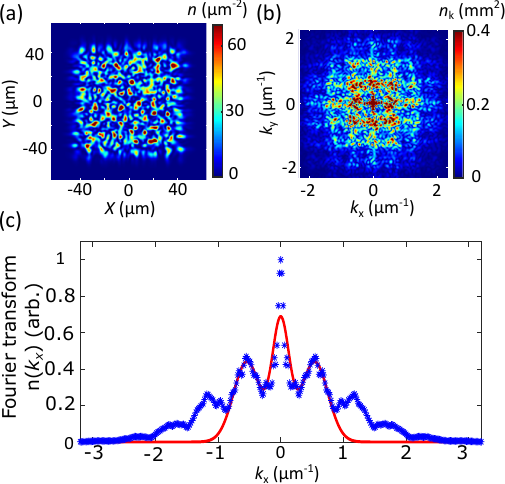}
\caption{Determination of the density profile periodicity. (a) Simulated 2D density in real space of BEC sample after 20~ms TOF. (b) Fourier transform of the simulated 2D density from (a). (c) Integrated Fourier transform in the $k_x$ direction (blue dots). Fitting the central peak and two side peaks(red line) yields the density periodicity.}
\label{fig:Fig_S01}
\end{figure}

First, we perform Fourier transform on the density profile to get the atomic distribution in reciprocal space ( Figs.~\ref{fig:Fig_S01}a and ~\ref{fig:Fig_S01}b). Since the system is symmetrical in the $k_x$ and $k_y$ direction, we sum over the $k_y$ direction to get the integrated 1D density profile(Fig.~\ref{fig:Fig_S01}c) and extract the period in the $k_x$ direction.
\begin{figure*}[ht]
\center
\includegraphics[width=\textwidth]{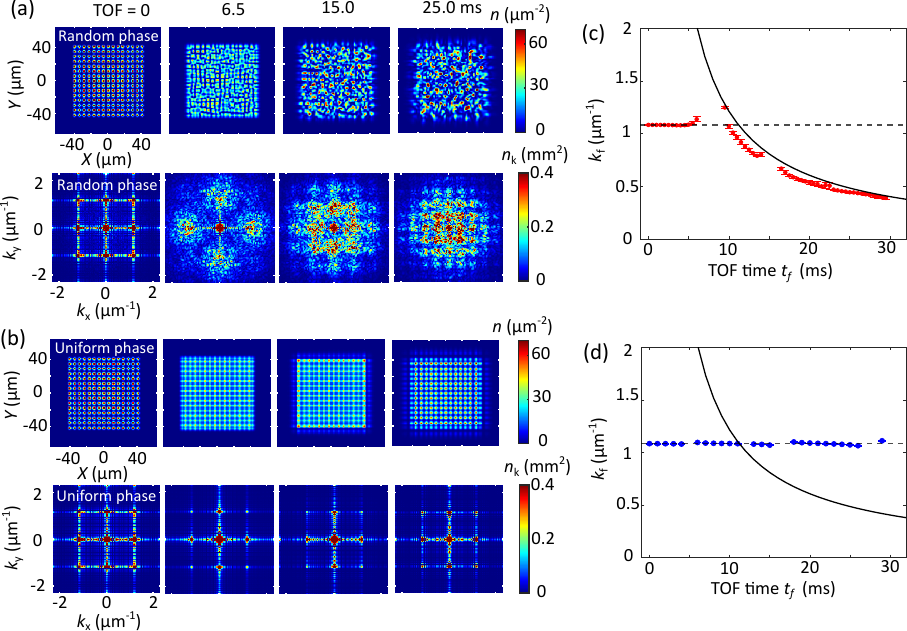}
\caption{Gross–Pitaevskii Equation(GPE) simulation of interference in a BEC array with random and constant relative phases. (a) Time evolution of a 2D array of BECs in free space, where the phases of the BECs in different sites are random. (b) Time evolution of a 2D array of BECs in free space where the phases of the BECs are constant. (c) The wave vector $k_f$ of the main side peaks in the Fourier spectrum (see main text). Red circles correspond to BECs with random phases, following $k_f=md/\hbar t_f$ (solid line). (d) Blue circles correspond to BECs with constant phases, following the initial BEC separation (dashed line).}
\label{fig:Fig_S0}
\end{figure*}

We extract the density periodicity from the positions of the dominant side-peaks in the momentum space. We fit the central peak and the two side peaks simultaneously with a sum of three Gaussians to obtain the positions of the side peaks at $k_x = \pm k_f$ (Fig.~\ref{fig:Fig_S01}c).

We use GPELab\cite{antoine2014gpelab,antoine2015gpelab} to simulate the experiment based on the Gross–Pitaevskii equation(GPE). In order to speed up the calculation, instead of performing 3D simulation, we assume the the system is a 2D gas confined in th x-y plane. In the z direction, the system is in the ground state of a harmonic trap with a $2\pi\times400$~Hz trap frequency. We use the simulation scheme $Relaxation$ in Ref.~\cite{antoine2015gpelab} to evolve the system from  $t_s$~=~0 to 30~ms. The time step is 0.005~ms. From $t_s$ = 0~ms, the sample is released into free space. The spatial range in both the x and y direction is -160 to 160~$\mu$m, and the size of the real space grid is $513\times513$. The initial wave function is a $15\times15$ BEC array with each site containing 800 atoms. The distance between neighboring sites is $d$~=~5.8~$\mu$m. The phase of different sites are set to be random for Fig.~\ref{fig:Fig_S0}a and uniform for Fig.~\ref{fig:Fig_S0}b. The more sites in an array, the more obvious is the random phase interference, which is consistent with the result in 1D arrays \cite{hadzibabic2004interference}.

The time-evolution results are shown in Fig.~\ref{fig:Fig_S0}. 
If the initial relative phases are random, an interference pattern appears after a few ms of evolution, see Fig.~\ref{fig:Fig_S0}a. The periodicity of the interference pattern is more clear in reciprocal space, see the bottom row of Fig.~\ref{fig:Fig_S0}a. We obtain the periodicity in Fourier space. The simulated $k_f$ for random phases roughly follow the solid line (Fig.~\ref{fig:Fig_S0}c) given by $\lambda=\hbar t_f/md$, where $\lambda = 2\pi/k_f$. This simulation result is consistent with our experiment observation.

On the other hand, if the initial phases of BECs are uniform, the periodicity remains constant over an evolution time of 30~ms (Fig.~\ref{fig:Fig_S0}b). This result suggests that, the period of the density is given by the initial period for a long time. The simulated $k_f$ for uniform phases is not consistent with our experimental observation.
Therefore, we conclude that the BEC samples in our experiment have random phases.

\subsection*{C. Determination of the Boundary between the Stable and Collapsed BEC}
\begin{figure}[ht]
\centering
\includegraphics[width=0.48\textwidth]{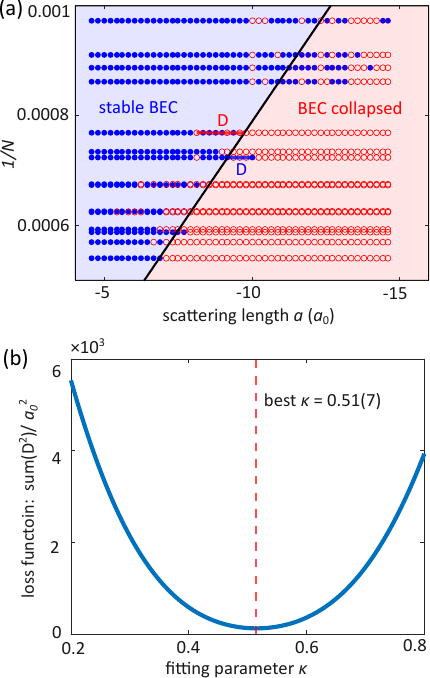}
\caption{Determination of the fitting constant $\kappa$ (a) When a data point falls on the wrong side of the boundary, see Eq.\ref{lineKappa}, such as red circles on the left of the solid line and blue circles on the right of the solid line, we calculate the distance $D$ from such a point to the boundary. The loss function is determined by summing over all these distances. (b) How the loss function changes with $\kappa$. The loss function reaches its minimum when $\kappa$ = 0.51(7), as indicated by the dashed line.}
\label{fig:Fig_S2}
\end{figure}

On the boundary of non-collapsed and collapsed samples, the reciprocal of the atom number $1/N$ is expected to be proportional to the negative scattering length $a_c$. Our task is to determine the ratio $\kappa$, such that 
\begin{equation}
a_c/\ell= -\kappa/N ,
\label{lineKappa}\end{equation}
where $a_c$ represents the critical scattering length of the BEC collapse with atom number $N$, $\ell=\sqrt{\hbar/m\bar{\omega}}$ is the harmonic length of the trap, and $\bar{\omega}$ denotes the geometric mean of the trap frequencies.

\begin{figure}[h!]
\centering
\includegraphics[width=0.48\textwidth]{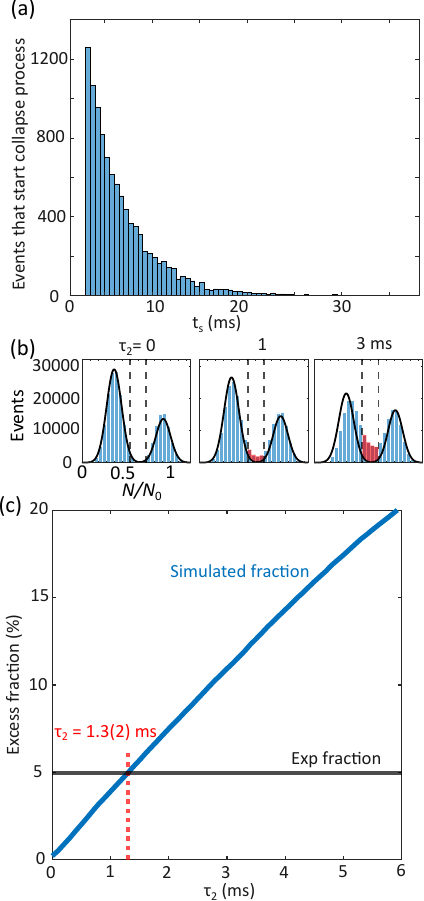}
\caption{Determination of $\tau_2$ by simulation.(a)Events that start the collapse process. (b) Histogram of the atom numbers from simulations. Black lines are fits based on the stable and collapsed BEC distributions. The red parts of the histogram represent excess events of BECs with intermediate atom numbers 0.6$<\bar{n}<$0.8. (c)Fraction of excess events. The blue line represents how the residual fraction change with $\tau_2$. The black line indicates the position of the residual fraction from our experiment,while the red dashed line represents the corresponding $\tau_2$ = 1.3(2)~ms.}
\label{fig:Fig_S1}
\end{figure}

Based on the experiment data shown in Fig.~\ref{fig:collapse}, we use least squares method to determine the optimal value of $\kappa$. The first step is to define a loss function. Given a parameter $\kappa$, we could draw a line $l_\kappa$. We assume that the samples to the left of the line $l_\kappa$ correspond to non-collapsed samples (depicted as blue dots in Fig.~\ref{fig:Fig_S2}a) and to the right to be collapsed samples (depicted as red circles in Fig.~\ref{fig:Fig_S2}a). If a data point is located on the wrong side, such as a red circle falling on the left side, we consider it as an error point. We measure the horizontal distance $D$ (as depicted in Fig.~\ref{fig:Fig_S2}a) from the error points to the line $l_\kappa$, and compute the sum of squares of $D$ as the loss. We plot the loss as a function of $\kappa$ (Fig.~\ref{fig:Fig_S2}b), and determine the optimal $\kappa$ value that minimizes the loss function. Our result of $\kappa = 0.51(7)$ is reported in the main text.

\subsection*{D. Stochastic Model of BEC Collapse and Simulations to Determine $\tau_2$}
As described in the main text, our stochastic model is composed of two steps. In the first step, while the instability increases, the atom number remains constant for a time $t_s$, which we treat as a random variable. In the second step, the BEC begins to collapse, leading to an exponential decrease in the atom number with a time constant $\tau_2$. 

From the experiment, we know that the probability distribution function(PDF) of the random variable $t_s$ consists of two parts. Initially, as the modulation instability requires time $t_D$ to grow up \cite{nguyen2017formation}, for $t_s < t_D$=3~ms, the probability of collapse is zero. Subsequently, after $t_D$, the probability of collapse would decrease with a time constant of $\tau_1 = 3.8$~ms. Utilizing this PDF, We first generate 10000 samples of $t_s$(Fig.~\ref{fig:Fig_S1}a).

For the simplicity in the simulation, we use normalized atom numbers. The initial atom number $N_0$ is set to 1. After BEC fully collapsed, there are still 40$\%$ of atoms remain. For a given sample, after a time evolution $t_h$, if $t_h \leq t_s$, the atom number $N_t$ remains unchanged. If $t_h> t_s$, $N_t$ would decay to $e^{-(t_h - t_s)}\times(1-0.4)+0.4$. Additionally, beyond this idealized process, there is fluctuation in atom number due to imperfect of loading. We model this fluctuation as a Gaussian distribution with $\sigma = 0.09$. This fluctuation is then added to the atom number to obtain the final atom number $N_t$.
We take $t_h$ to be 0~ms to 20~ms with a spacing of 1~ms. Subsequently, we obtain the histogram of all events at all time $t_h$(Fig.~\ref{fig:Fig_S1}b).

From the experiment, we observe that the atom number of non-collapsed and collapsed samples follow Gaussian distributions with $\sigma = 0.09$ and centered at 1 and 0.4 respectively, as indicated by red lines in Fig.~\ref{fig:Fig_3}b. Therefore, we fit the histogram with Gaussian distributions representing non-collapsed and collapsed samples. There would be excess part, particularly between 0.6 to 0.8 (Fig.~\ref{fig:Fig_S1}b, Figs.~\ref{fig:Fig_3}e and f). We consider  the residual part as samples in the middle of the process of losing particles and calculate the fraction of residual as a function of $\tau_2$(Fig.~\ref{fig:Fig_S1}c). In the experiment, this fraction is measured to be 5$\%$. Then, we could determine the optimal $\tau_2$ that matches this fraction to be 1.3~ms.

\end{document}